\documentclass[floatfix,twocolumn,english,aps,arxiv,superscriptaddress]{revtex4-1}
\usepackage{grffile}
\RequirePackage{fix-cm} 
\DeclareMathSizes{10}{10}{5}{5}
\usepackage{gensymb}
\usepackage[latin9]{inputenc}
\usepackage{color}
\usepackage{babel}
\usepackage{amsmath}
\usepackage{amssymb}
\usepackage{amsfonts}
\usepackage{graphicx}
\usepackage{epstopdf}
\usepackage{multirow}
\usepackage{siunitx}
\usepackage{cancel}
\PassOptionsToPackage{normalem}{ulem}
\usepackage{ulem}
\usepackage[version=3]{mhchem}
\usepackage[unicode=true, bookmarks=false, breaklinks=false,pdfborder={0 0 1},colorlinks=false] {hyperref}
\usepackage{subcaption}
\begin{document}

\title{Fermiology and transport properties of the candidate topological crystalline insulator SrAg$_4$Sb$_2$}

\author{J Green}
\affiliation{Department of Physics and Astronomy and California NanoSystems Institute, University of California, Los Angeles,
CA 90095, USA}

\author{Eve Emmanouilidou}
\affiliation{Department of Physics and Astronomy and California NanoSystems Institute, University of California, Los Angeles,
CA 90095, USA}

\author{Harry W. T. Morgan}
\affiliation{Department of Chemistry and Biochemistry, University of California, Los Angeles,
CA 90095, USA}

\author{William T. Laderer}
\affiliation{Department of Chemistry and Biochemistry, University of California, Los Angeles,
CA 90095, USA}

\author{Chaowei Hu}
\affiliation{Department of Physics and Astronomy and California NanoSystems Institute, University of California, Los Angeles,
CA 90095, USA}

\author{Jonathan Loera}
\affiliation{Department of Physics and Astronomy and California NanoSystems Institute, University of California, Los Angeles,
CA 90095, USA}

\author{Anastassia N. Alexandrova}
\affiliation{Department of Chemistry and Biochemistry, University of California, Los Angeles,
CA 90095, USA}

\author{Ni Ni}
\email{Corresponding author: nini@physics.ucla.edu}
\affiliation {Department of Physics and Astronomy and California NanoSystems Institute, University of California, Los Angeles, CA 90095, USA}
\begin{abstract}

Compared to time-reversal symmetry-protected $\mathbb{Z}_2$ topological insulators and Dirac/Weyl semimetals, there are significantly fewer candidates for topological crystalline insulators. SrAg$_4$Sb$_2$ is predicted to exhibit topological crystalline insulator behavior when considering spin-orbit coupling. In this study, we systematically investigate single crystals of \ce{SrAg4Sb2} using electrical transport and magnetic torque measurements, along with first-principles calculations. Our transport data reveals its compensated semimetal nature with a magnetoresistance up to around 700\% at 2 K and 9 T. Analysis of de Haas-van Alphen oscillations uncovers a Fermi surface consisting of three distinct Fermi pockets with light effective masses. Comparison between the three-dimensional fermiology obtained from our oscillation data and the first-principles calculations demonstrates excellent agreement. This confirms the accuracy of the calculations, which indicate a band inversion centered at the $\Gamma$ point and identify the existence of nontrivial tube and needle hole Fermi pockets at $\Gamma$, alongside one trivial diamond electron pocket at the T point in the Brillouin zone. Furthermore, symmetry and topology analysis results in two potential sets of topological invariants, suggesting the emergence of two-dimensional gapless Dirac surface states either on the $ab$ planes or on both the $ab$ planes and mirror planes, protected by crystal symmetries. Therefore, SrAg$_4$Sb$_2$ emerges as a promising candidate topological crystalline insulator.

\end{abstract}
\date{\today}
\maketitle

\section{Introduction}

In the past decade, topological crystalline insulators (TCIs) have attracted interest due to their robust topological surface states on  certain high symmetry crystal surfaces. Unlike the well-known $\mathbb{Z}_2$ topological insulators protected by time reversal symmetry, these topological states are protected by crystal symmetries \cite{TCI, TCIinvariants}. This unique property of TCIs makes their topological property more robust against external perturbations such as magnetic fields and readily tunable through the application of strain, structural distortions, and more\cite{hsiehsnte,tanaka-3,serbyn, evelyn, spinfiltered, nanomembranes}, providing a versatile application platform in spintronics, quantum computing, and pressure sensors \cite{applications}. The first TCI, SnTe, was predicted in 2012 and has since been studied extensively \cite{hsiehsnte,tanaka,tanaka-2,GiantMR, Transport-1, Transport-2, Transport-3, ARPESpump, ARPES-1, ARPES-2, ARPES-3}. However, compared to $\mathbb{Z}_2$ topological insulators and Dirac/Weyl semimetals, the exploration of TCI insulator candidates is far more limited. 

Guided by databases \cite{tqc, catalog, catalog-2}, we found that according to first-principles calculations, \ce{SrAg4Sb2} can be classified as a TCI when spin-orbit coupling (SOC) is taken into account. \ce{SrAg4Sb2} crystallizes in the \ce{CaCu4P2}-type centrosymmetric trigonal space group $R\bar{3}m$ (No.166) with \ce{Sr} atoms sandwiched in between layers of \ce{Ag2Sb}, as shown in the right inset of Fig. \ref{xray}. In a recent study of the electronic structure and topology of \ce{SrAg4Sb2}, density functional theory (DFT) calculations were performed both in the presence and absence of SOC \cite{morgan_SrAg4Sb2_bonding}. In \ce{SrAg4Sb2}, one might expect SOC to play an important role Since SOC is most significant when heavy atoms are present. According to the calculated band structures, there are two band touching points at the $\Gamma$ and T points of the Brillouin zone that are gapped by SOC with band gaps of $\approx$ 0.5 eV. More specifically, the two bands immediately above the Fermi level exchange character between \{Ag $s$ + Sb $s$\} and \{Ag ${d_{xy}, d_{x^2-y^2}}$ + Sb ${p_x, p_y}$\} as a function of $k_z$. It is also worth noting that this is a relatively ``clean'' band structure \cite{catalog-2}, making \ce{SrAg4Sb2} a great candidate TCI. 

However, despite reports of the observation of quantum oscillations (QOs) \cite{Sr142}, there is no thorough experimental study of this compound to determine its three-dimensional (3D) fermiology to shed light on if it is indeed a TCI or not. In this paper, we report the single crystal growth, magnetotransport properties, 3D fermiology extracted from angle-dependent QOs, as well as density functional theory (DFT) calculations for \ce{SrAg4Sb2}. By magnetotransport measurements, we show that \ce{SrAg4Sb2} is a compensated semimetal. Through the data analysis of the angle-dependent de Haas-van Alphen (dHvA) effect in magnetic torque measurements and the comparison with theoretical calculations, we are able to verify the existence of two hole pockets and one electron pocket. The excellent agreement between DFT and experiments suggests that our DFT calculations as well as the ones performed in databases \cite{tqc, catalog, catalog-2} correctly describe the band structure of \ce{SrAg4Sb2}. We further show that a band inversion centered around the T point with an avoided band crossing along the $\Gamma$-T line exists on the mirror planes and then discuss the potential topological invariants and where the 2D Dirac surface states protected by crystal symmetries are expected. Our result provides evidence that \ce{SrAg4Sb2} is a promising TCI for exploration of topological surface states protected by crystal symmetry.

\section{Experimental Methods}

\begin{figure}
    \centering
    \includegraphics[width=\columnwidth]{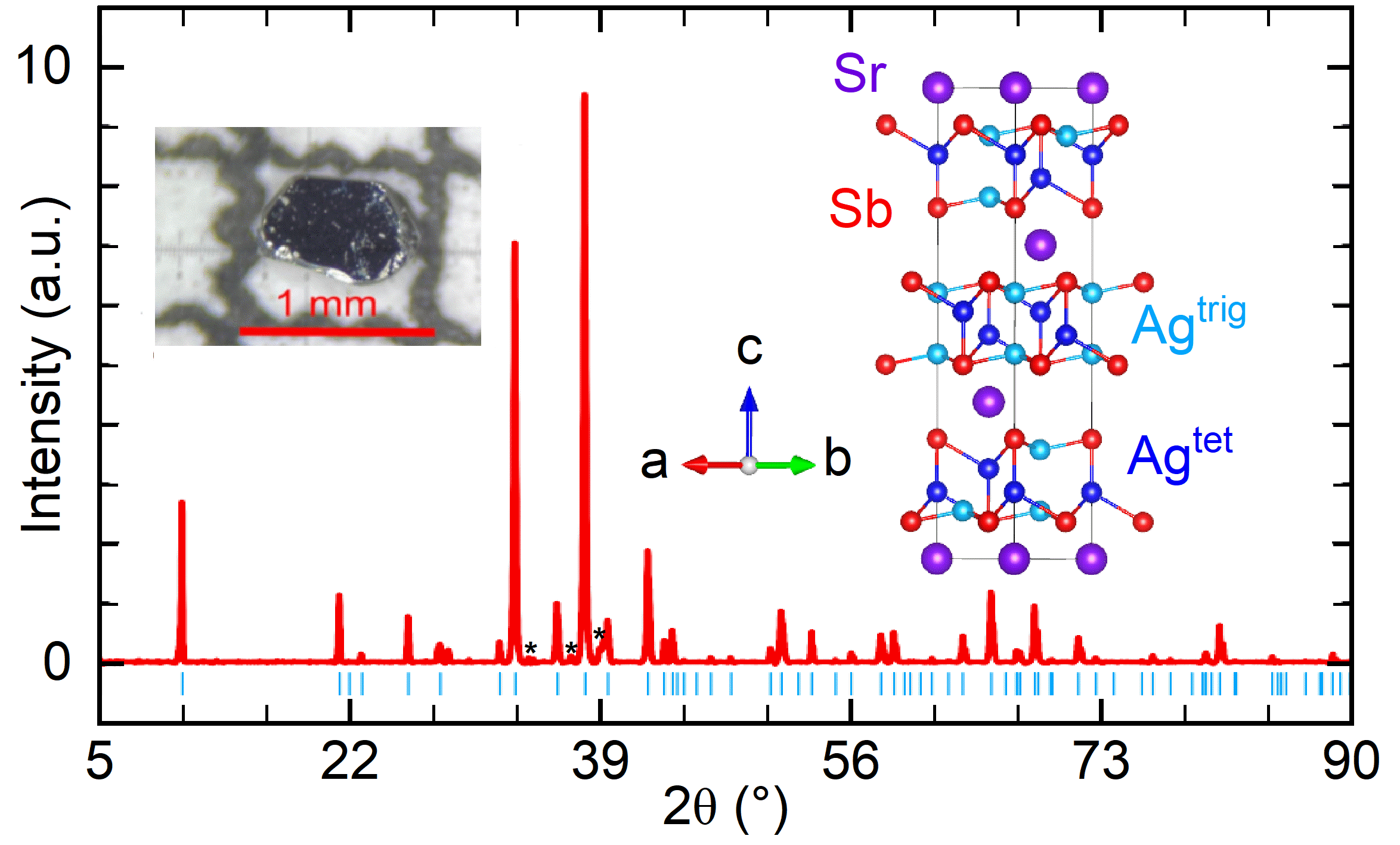}
    \caption{The powder X-ray diffraction pattern of crushed \ce{SrAg4Sb2} single crystals. Peaks can be indexed using the experimentally determined crystal structure (blue tick marks) \cite{stoyko}. The asterisks indicate peaks from the small amount of \ce{Ag3Sb} that was present on the crystals. Left inset: A photo of a typical crystal against a 1-mm sized grid. Right inset: The crystal structure of \ce{SrAg4Sb2}. Ag$^{\rm{tet}}$: the Ag atoms connected to 4 Sb atoms in tetrahedral coordination. Ag$^{\rm{trig}}$: the trigonally coordinated Ag atoms.}
    \label{xray}
\end{figure}

Single crystals were grown using the self-flux method \cite{srag4as2, euag4as2}. Sr, Ag and Sb pieces were combined using the molar ratio 1:14:7 and placed inside an alumina crucible, which was then sealed in a quartz tube under vacuum. When the growth ampule was heated to 1050 $\celsius$ overnight for the mixture to homogenize and then allowed to slowly cool down to 650 $\celsius$, the growth of \ce{SrAg4Sb2} single crystals was complicated by the existence of another stable, competing phase, \ce{SrAgSb}. Where \ce{SrAgSb} was sandwiched between \ce{SrAg4Sb2}, suggesting that SrAgSb formed first. This was confirmed by performing another trial with an increased spin-out temperature of 750 $\celsius$, where the relative amount of \ce{SrAgSb} was higher. Thus, to eliminate the presence of \ce{SrAgSb}, after heating the quartz tube to 1050 $\celsius$, we quenched it in air. It was then placed into a furnace which was at 700 $\celsius$, and then slowly (over 120 hours) cooled to 580 $\celsius$, at which point it was centrifuged to separate the single crystals from the liquid flux. By this, the presence of \ce{SrAgSb} was completely eliminated, as illustrated by Fig. \ref{xray}. Free standing three dimensional, hexagonal single crystals of \ce{SrAg4Sb2} formed, as shown in the inset of Fig. \ref{xray}. 

Powder X-ray diffraction measurements were performed using a PANalytical Empyrean (Cu K$\alpha$ radiation) diffractometer. Magnetotransport measurements were performed inside a Quantum Design (QD) Dynacool Physical Properties Measurement System (PPMS) with a maximum magnetic field of 9 T. Magnetic torque measurements were performed in a QD-PPMS. Torque measurements were made by mounting a small piece of single crystal on the tip of a piezoresistive cantilever. The magnetic torque was then inferred from the magnetoresistance of the cantilever measured by a Wheatstone bridge, as the resistance of the cantilever is very sensitive to the deformation caused by torque. The electrical resistivity ($\rho_{xx}$) and Hall ($\rho_{yx}$) measurements were performed using the six-probe technique inside QD-PPMS. To eliminate unwanted contributions from mixed transport channels, data were collected while sweeping the magnetic field from -9 T to 9 T. The data were then symmetrized to obtain $\rho_{xx}(B)$ using $\rho_{xx}(B)=(\rho_{xx}(B)+\rho_{xx}(-B))/2$ and antisymmetrized to get $\rho_{yx}(B)$ using $\rho_{yx}(B)=(\rho_{yx}(B)-\rho_{yx}(-B))/2$. The magnetoresistance is defined as $\mathrm{MR} = (\rho_{xx}(B)-\rho_{xx}(0))/\rho_{xx}(0)$. In our measurement geometry, a positive slope of $\rho_{yx}(B)$ suggests the hole carriers dominate the charge transport. 

The electronic structure of \ce{SrAg4Sb2} was studied via DFT calculations using the PBE functional\cite{hohenberg,kohn,pbefunctional} and the projector augmented wave (PAW) pseudopotential method as implemented in the Vienna Ab initio Simulation Package (VASP), version 5.4.4 \cite{blo}.
An approximate SOC correction implemented in VASP was used to compute the electronic properties in the first Brillouin zone \cite{vasp soc}.
Fermi surfaces were computed using a $k$-mesh spacing of 0.008 (31-31-31).
Fermi surface data were generated from the DFT calculations using Vaspkit and visualized with Fermisurfer \cite{vaspkit,fermisurfer}.
Band structures were plotted with pyprocar version 5.6.6 \cite{pyprocar}.
dHvA frequencies were computed using SKEAF, with a modified file conversion program to allow SKEAF to process VASP output \cite{skeaf}.

\begin{figure}
\centering
    \includegraphics[width=\columnwidth]{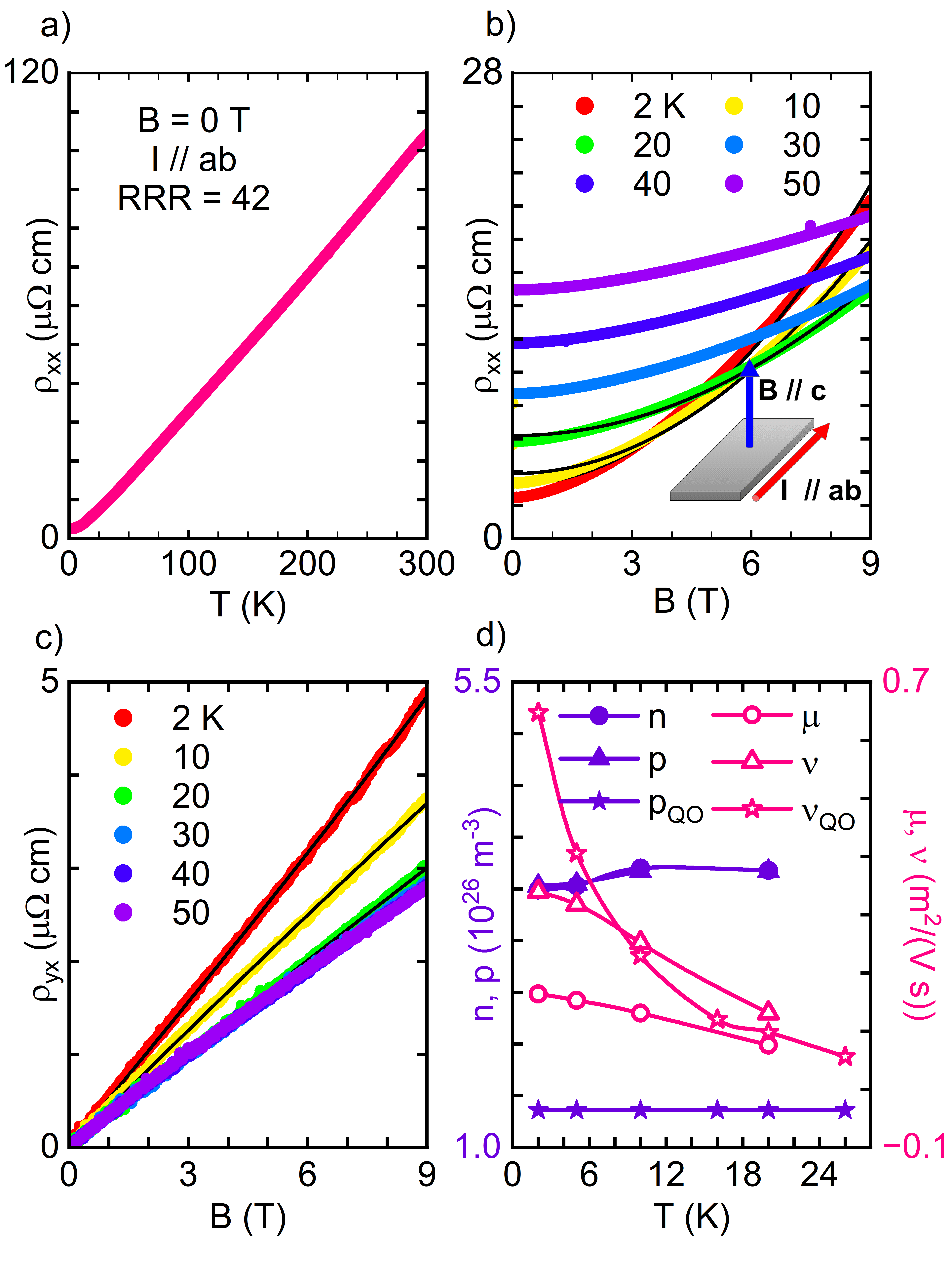}
    \caption{(a) The temperature-dependent resistivity of \ce{SrAg4Sb2}. (b), (c) The field-dependent resistivity $\rho_{xx}$ and Hall resistivity $\rho_{yx}$ measured at temperatures between 2 K and 50 K. Filled circles represent experimental data collected, while the solid black lines are generated form the two-band model fit, see text. (d) The carrier concentrations and mobilities. Two methods are used to obtain the values. One is through the two-band model fit of $\rho_{xx}$ and  $\rho_{yx}$ up to 20 K; the other is by analyzing the QO data measured with the field along the [001] direction.}
    \label{transport}
\end{figure}

\section{Experimental Results and Discussion}

\ce{SrAg4Sb2} has the lattice parameters $a$ = $b$ = 4.7404(4) \AA, $c$ = 25.029(2) \AA, $\alpha$ = $\beta$ = 90$^o$ and $\gamma$ = 120$^o$ \cite{stoyko, Ca142}. Figure \ref{xray} shows the powder X-ray diffraction pattern where all peaks except three (marked by an asterisk) can be well indexed using the \ce{SrAg4Sb2} structure. These three anomaly peaks are likely due to small amount of \ce{Ag3Sb} droplets present on the surfaces of the crystals. 

Figure \ref{transport}(a) shows the temperature dependence of the resistivity of \ce{SrAg4Sb2} measured at zero field and with the current in the $ab$ plane. The residual resistivity ratio (RRR), defined as $\rho_{300K}/\rho_{2K}$ is 42, and the residual resistivity is \SI{2.5} {\micro\ohm ~cm}. Upon cooling, the resistivity decreases with a linear behavior in the range between 300 K and 20 K, a characteristic feature of a conventional metal. Below 20 K, the resistivity exhibits the typically Fermi-liquid behavior $\rho = \rho_0 + \alpha T^\beta$, where $\beta$ = 2 \cite{FermiLiquid}. 
\subsection{Magnetotransport Properties of \ce{SrAg4Sb2}}
Figures \ref{transport}(b) and 2(c) show the magnetoresistivity $\rho_{xx}$ and Hall resistivity $\rho_{yx}$ respectively with $B\parallel c$ and $I\parallel ab$ at temperatures in the range between 2 K - 50 K. $\rho_{xx}$ shows a nearly parabolic behavior with a moderately large MR of 700\% at 2 K. Quasi-linear $\rho_{yx}$ shows QOs in a field range of 3 T to 9 T. Large quadratic MR with quasi-linear $\rho_{yx}$ are the characteristic features of compensated semimetals. To extract the carrier concentrations and mobilities, the data were fitted using the semiclassical two-band Drude model of transport \cite{pippard}. The field dependence of $\rho_{xx}$ and $\rho_{yx}$ are given by  

 \begin{equation}
 \rho_{xx} = E_x/J_x=\frac{n \mu + p\nu + (n\nu +p\mu)\mu\nu B^2}{e(n\mu + p\nu)^2 + e(p-n)^2\mu^2\nu^2B^2}
 \end{equation}
 and
 \begin{equation}
\rho_{yx} = E_y/J_x=\frac{B(p\nu^2 - n\mu^2) + (p - n)\mu^2\nu^2B^3}{e(n\mu + p\nu)^2 + e(p-n)^2\mu^2\nu^2B^2},
 \end{equation}
 where $n$, $p$, $\mu$ and $\nu$ are the carrier densities and mobilities of electrons and holes respectively. The simultaneous nonlinear least-squares fit of $\rho_{xx}$ and $\rho_{yx}$ at 2 K using the above expressions yields $n = 3.49(5) \times 10^{26}$/m$^{3}$ and $p = 3.53(5) \times 10^{26}$/m$^{3}$. The ratio ${p}/{n}$ being approximately equal to 1 suggests that this is a compensated semimetal and consistent with the observation of large MR at low temperatures. The obtained electron and hole mobilities at 2 K are $\mu = 0.19(3)$ m$^2/(Vs)$ and $\nu = 0.35(6)$ m$^2/(Vs)$. As shown in Fig. \ref{transport} (d), upon increasing temperatures, the mobilities decreases due to the enhanced thermal fluctuations while the carrier densities remains unchanged. These carrier densities and mobilities are consistent with the ones we obtained from QO data which will be discussed in the following section.

\subsection{Quantum Oscillations of \ce{SrAg4Sb2}}

\begin{table*}
\caption{ Parameters extracted from dHvA data at three different field orientations. $\alpha$: $40\degree$ from [100], as shown in the inset of Fig. 3(a). Errors associated with frequencies F were found via the Full Width Half Maximum technique. While errors for $m^*_{exp}$ and $T_D$ were found by computing the variance-covariance matrix from the fitting function produced from fitting the data with the Levenberg-Marquardt (L-M) Algorithm in Origin. All other errors were found by way of error propagation.}
\label{meff}
\centering
\begin{tabular}{ c| c| c | c | c | c | c | c|  c | c|c }
 \hline
 \hline
           &$B$ & $F$(T) & $m^*_{\rm{exp}} (m_e)$  & $m^*_{\rm{DFT}} (m_e)$  & $k_F$($\si{\AA^{-1}}$) & $v_F$ ($10^{5}$m/s) & $T_D$(K) & $\tau_Q$($10^{-13}$s) & $\nu_Q$ ($\si{m^2/Vs}$) & ${p_{\rm{QO}}(10^{26}/\si{m^{3}})}$  \\
\hline
 \multirow{3}{*}{$F_{N}$}  & $\parallel \alpha$& 98(4)  & 0.14(1) & 0.06  & 0.06(1) & 4.5(8) & 3.6(1) & 3.4(1) & 0.43(3) &\multirow{3}{*}{0.06(2)}  \\
          &$\parallel [120]$& 340(9) & 0.35(2) & 0.22  & 0.10(2)  & 3.3(2) & 5.7(1) & 2.10(4) & 0.10(1)  \\  
          &$\parallel [001]$& 60(3)  & 0.097(1) & 0.04  & 0.04(1) & 5(1) & 2.4(1) & 5.0(2) & 0.63(7)   \\
 \hline
  \multirow{2}{*}{$F_{T1}$}  & $\parallel \alpha$ & 850(7)  & 0.35(4) & 0.39 & 0.16(1)  & 5.3(7) &     &     &   &\multirow{4}{*}{1.3(3)}    \\
          &$\parallel [001]$& 420(8)  & 0.27(1) & 0.25 & 0.11(2)  & 4.7(9) &     &     &       \\
   \cline{1-10}     
 \multirow{2}{*}{$F_{T2}$} &$\parallel \alpha$& 1120(9) & 0.43(4) & 0.42 & 0.18(2)  & 4.8(7) &     &     &       \\
          &$\parallel [001]$& 790(7)  & 0.23(2) & 0.33 & 0.16(1)  & 8.0(9) &     &     &       \\
   \hline     
 \multirow{2}{*}{$F_{D}$}  &$\parallel \alpha$ & 600(14)  & 0.16(1) & 0.26 & 0.14(2) & 9.5(2)  &     &     &      \\
          &$\parallel [120]$& 560(14)  & 0.23(2) & 0.23 & 0.13(2) & 5.4(2)  &     &     &      \\
\hline
\hline

\end{tabular}
\end{table*}

\begin{figure*}
    \includegraphics[width=\textwidth]{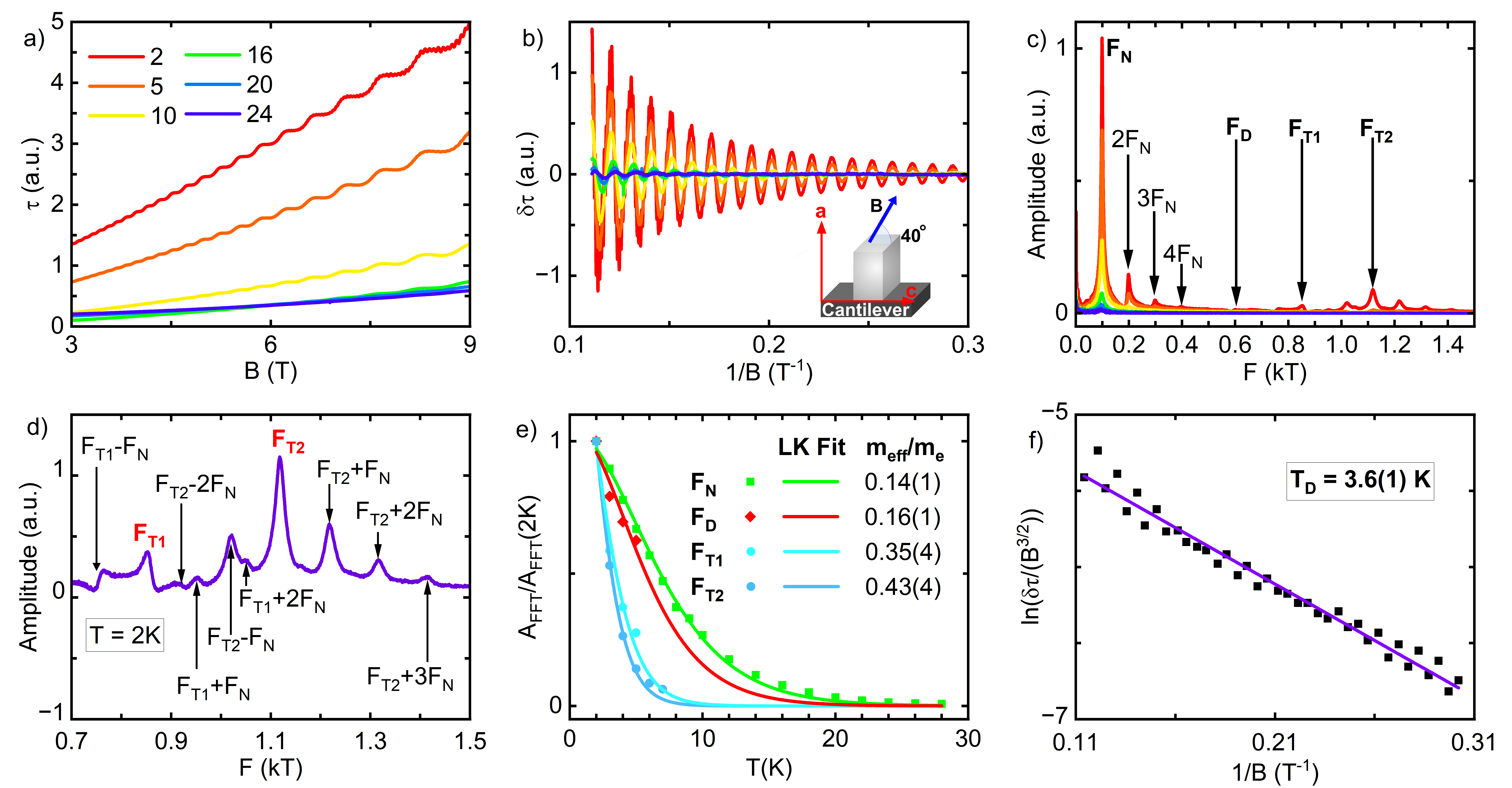}
    \caption{(a) $\tau$ at temperatures 2 - 24 K in applied fields ranging from 3 T to 9 T. (b) $\Delta \tau$ at the same temperatures as a function of 1/B. The bottom right inset contains the measurement configuration. (c) The FFT curves associated with (b). (d) The zoom-in plot of the peaks around F$_{T1}$ and F$_{T2}$. The surrounding peaks are due to magnetic interaction and the associated calculated frequencies are indicated by black arrows. (e) The normalized temperature-dependent FFT amplitude. The LK fits used to extract the effective masses are shown as the solid lines. (f) A Dingle Plot from QO data at 2 K, where $\ln({\Delta\tau/B^{3/2}})$ is plotted against $1/B$.} 
    \label{dHvA}
\end{figure*}

 \begin{figure*}
    \includegraphics[width=\textwidth]{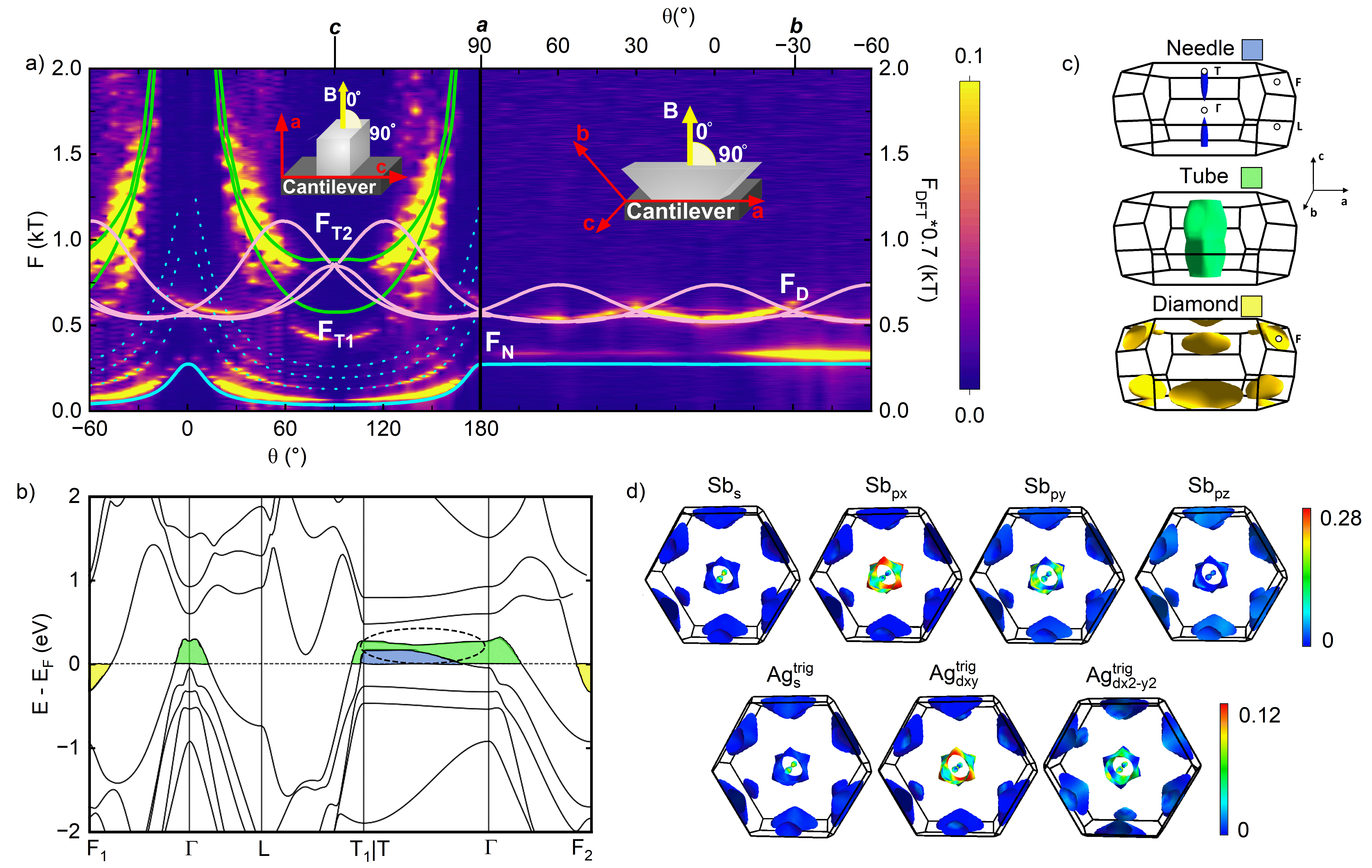}
    \caption{(a) Contour plot of the experimental FFT frequencies of $\Delta \tau$ as a function of angle. Overlaid are solid green, pink, and blue lines which indicate the DFT calculated frequencies associated with the tube, diamond, and needle pockets, respectively. The blue dashed lines are a guide for the eye to highlight the observed 2$^{nd}$, 3$^{rd}$, and 4$^{th}$ harmonics of F$_{N}$. In each panel, an inset with the measurement geometry can be found. (b) The DFT calculated electronic band structure. The dashed circle indicates the avoided band crossing along $\Gamma$-T line. (c) 3D renderings of each component of the Fermi surface are shown separately and annotated with relevant high-symmetry points, for visual clarity. The colored square next to each Fermi surface component corresponds to the colored band in (b). (d) Top row: 5$s$ and 5$p$ valence orbitals of Sb. Bottom row: Orbitals of \ce{Ag^{trig}} which contribute significantly to the topological band inversion - 5$s$, 4$d_{xy}$, and 4$d_{x^2-y^2}$.}
    \label{Fig4}
\end{figure*}

 \begin{figure*}
    \centering
    \includegraphics[width=0.6\textwidth]{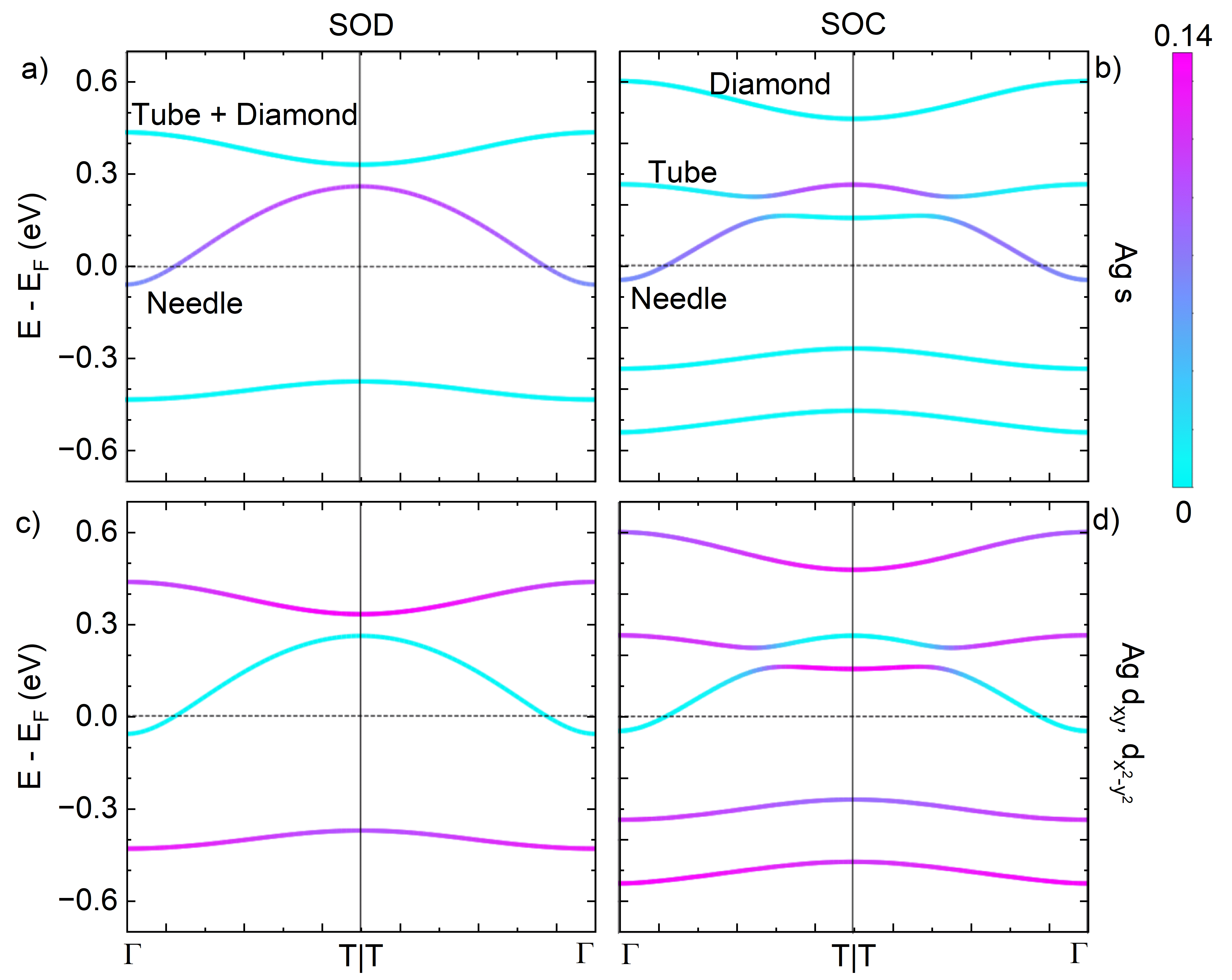}
    \caption{Band structures for the needle and tube bands along the $\Gamma$-T-$\Gamma$ $k$-path projected onto Ag$^{\rm{trig}}$ $s$, and Ag$^{\rm{trig}}$ $d_{xy} + d_{x^2-y^2}$. Panels (a) and (c) illustrate the SOD band structure and to the right of these panels, panels (b) and (d), show the SOC band structures.}
    \label{BandPMOs}
\end{figure*}

QOs are also observed in magnetic torque $\vec{\tau} = \vec{M} \times \vec{B}$ measurements. Figure \ref{dHvA}(a) depicts the field-dependent $\tau$ at various temperatures when the magnetic field was applied 40$\degree$ away from the $a$-axis, in the $ac$ plane, as shown in the inset of Fig. \ref{dHvA} (b). Strong QOs can be clearly seen above 3 T. Figure \ref{dHvA}(b) shows $\delta\tau$ after the subtraction of a polynomial background from $\tau$. The Fast Fourier Transform (FFT) of the oscillations (Fig. \ref{dHvA}(c)) reveal many features. One strong peak with the frequency $F_{N}$ = 98(4) T is observed, followed by three weaker peaks that are the 2nd, 3rd, and 4th higher harmonics of $F_{N}$, respectively. Peaks with $F_{D}$ = 600(14) T, $F_{T1}$ = 850(7) T, and $F_{T2}$ = 1120(9) T as well as lobes surrounding $F_{T1}$ and $F_{T2}$ also present. To better investigate this rich FFT feature, the zoom-in plot between 700 T and 1500 T at 2 K is plotted as Fig. \ref{dHvA}(d), where QO FFT peaks other than $F_{T1}$, $F_{T2}$ and $F_{N}$ are clearly seen. The black arrows point to the exact location of the calculated values of $F_{T1}$ $\pm$ $nF_{N}$ and $F_{T2}$ $\pm$ $nF_{N}$ ($n=1, 2, 3)$. The positions of these lobes agree well with the locations of the black arrows. Frequency combinations such as this can arise from  magnetic breakdown (MB) or magnetic interaction (MI). MB occurs at fields high enough such that the separation $\hbar\omega_c = e\hbar B_0/m^*$ between Landau levels is larger or equivalent to $E^2_g/E_F$ \cite{Blount}, where $E_g$ is the energy gap between the two bands and $E_F$ is the Fermi energy. An estimate of the magnetic breakdown field $B_0$ can be made through $B_0=(\pi\hbar/2e)(k_1-k_2)^3/(k_1+k_2)$ \cite{Chambers}, where $k_1$ and $k_2$ are the Fermi wave vectors of the Fermi pockets contributing to the combined oscillation frequencies, respectively. 
Plugging in values as shown later in this section, we find that the field needed for MB to occur between the Fermi pockets associated with $F_N$ and $F_{T1}$ is more than 400 T, suggesting this effect is a result of MI instead. MI can occur as a complication of sample shape and/or crystal anisotropy much like that seen in layered materials \cite{MagInt}. Figure \ref{dHvA}(e) summarizes the temperature-dependent FFT amplitudes of $F_N$, $F_D$, $F_{T1}$ and $F_{T2}$. The amplitude of multiple oscillations in the magnetic torque is given by the Lifshitz-Kosevich (LK) theory as

\begin{equation} 
    \Delta \tau (B) = \pm \Sigma_i A_i B^{3/2} R_T^i R_D^i R_S^i \: \textrm{sin}[2\pi (\frac{F_i}{B} + \phi_i)],
    \label{osc}
\end{equation}
where $A_i$, $F_i$ and $\phi_i$ are constants, frequencies and phase factors of each pocket, respectively\cite{schoenberg}. $R_T$ represents the thermal damping factor, which is a finite temperature correction to the Fermi-Dirac distribution. It describes the temperature dependence of the oscillations' amplitude and is given by equation $R_T = X/{\rm{sinh}}X$, where $X=\alpha T m^*/Bm_e$, $\alpha$ is equal to $14.69$ T/K and $m^*$ is the cyclotron effective mass. $R_D = {\rm{exp}}(-\alpha T_{D} m^* / Bm_e)$ is the Dingle damping factor, which is related to the quantum lifetime through the equation $\tau_q=\hbar/(2\pi k_BT_{D})$. $R_S$ = cos $(\pi g m^*/2m_e)$ is the spin damping factor, which accounts for the interference between two oscillations from spin-split Landau levels. When several frequencies exist and are not easily separable, the extraction of the effective mass and the Dingle temperature can be quite challenging or even impossible by fitting the QO using Eq. (3). Alternatively, we obtain the effective mass by fitting the temperature-dependent FFT amplitude of each peak $A_{\rm{FFT}}$ (Fig. \ref{dHvA}(e)) using $A_{\rm{FFT}} \propto X/ {\rm{sinh}}X$, where $B$ is the average inverse field of the FFT window from $B_1$ to $B_2$ and defined as $1/B = (1/B_1 + 1/B_2)/2$. However, care must be taken when choosing which $B_1$ and $B_2$ to use in the analysis, as the wrong choice may lead to overestimated or underestimated values. For reasons outlined in our previous report on \ce{SrAg4As2}, we chose $B_1=5$ T and $B_2=9$ T\cite{srag4as2}. The obtained effective masses are $m^*_{N}=0.14(1)m_e$, $m^*_{D}=0.16(1)m_e$, $m^*_{{T1}}=0.35(4)m_e$ and $m^*_{{T2}}=0.43(4)m_e$. Despite there being many frequencies present, the $F_N$ peak dominates the oscillation, resulting in the exponential decaying oscillation amplitude with increasing $B$ as shown in Fig. \ref{dHvA}(b). The maximum peak intensity at 2 K in Fig. \ref{dHvA}(b), $\Delta \tau _{\rm{max}}$, as well as the corresponding field $B$ are recorded and plotted in Fig. \ref{dHvA} (f). By fitting this Dingle plot with $\Delta \tau _{\rm{max}}/B^{3/2} \propto {\rm{exp}}(-\alpha T_{D} m^*_{N} / Bm_e)$ where $T_{D}$ is the fitting parameter, we are able to extract a Dingle temperature of $T_{D}$(2 K) = 3.6(1) K. From this we calculate the quantum lifetime to be $\tau_q$ (2 K) = $3.4(1)\times 10^{-13}$ s. Using $\tau_q$ = ${m^*\mu_q}/{e}$ we estimate the quantum mobility associated with the $F_N$ pocket to be $\mu_q=0.43(3)$ m$^2/\rm{(Vs)}$. Similar analysis is performed for the QO data at various temperatures and with the field along the [120] and [001] directions. The obtained effective masses alongside the DFT computed ones are summarized in table \ref{meff}. The obtained $m^*$s are angle-dependent, which is expected for anisotropic Fermi pockets. 
$\mu_q$ ranges from 0.63(7) m$^2/\rm{(Vs)}$ when holes move in the $ab$ plane to 0.10(1) m$^2/\rm{(Vs)}$ when holes move perpendicular to the [120] direction, as expected for a compound with layered structure. The exacted quantum mobility associated with the $F_N$ pocket when the field is along the [001] direction monotonically decreases upon warming, as plotted in Fig. 2(d). The hole mobility extracted using the two-band model fitting, 0.35(6) m$^2/\rm{(Vs)}$, is an average of the classical mobilities for all hole pockets. It is comparable to the ones obtained from QO, indicating that the impact of small angle scattering is likely negligible \cite{Cd3As2}.

Since QO frequency is directly proportional to the extreme cross sectional area ($S$) of the Fermi surface perpendicular to the magnetic field through the Onsager relation $F = (\hbar/2\pi e) S$ \cite{schoenberg}, to reconstruct the Fermi pockets, we utilized angle-dependent magnetic torque measurements to obtain the angular dependence of the extreme cross sectional area of each Fermi pocket. The sample was rotated from $a$ to $c$ in the $ac$ plane, as well as in the $ab$ plane to the crystallographic $a$ axis. Rotation in these two planes allows for the experimental determination of the 3D fermiology of the material. A contour plot of the frequencies extracted via FFT of $\delta\tau$ can be seen in Fig. \ref{Fig4}(a). The left panel illustrates rotation in the $ac$ plane where four branches of the fundamental dHvA frequencies ($F_{N}$, $F_{T1}$, $F_{T2}$, and $F_{D}$) and a set of three harmonic frequencies of $F_{N}$ are observed (dashed lines are a guide for the eye). In the right panel where the rotation is in the $ab$ plane, only two fundamental frequencies ($F_{N}$ and $F_{D}$) are observed with $F_{N}$ being angle independent.

\subsection{Comparison to DFT calculations}

To understand the angular dependence of QO, DFT calculations were performed and the results were compared to the measured dHvA data.
We first optimized the structure of \ce{SrAg4Sb2} and found good agreement with experiment\cite{stoyko,morgan_SrAg4Sb2_bonding}.
We then computed the band structure and Fermi surface, shown in Fig. \ref{Fig4}(b) and (c). 
The computed Fermi surface contains three features: the needle pocket, a long, closed hole pocket oriented along $z$; the tube pocket, a continuous distorted cylinder oriented along $z$ which encloses the needle pocket; and the diamond pocket, a set of three rounded-diamond-shaped electron pockets centered at the F points on the faces of the 1st Brillouin zone.
Solid lines on Fig. \ref{Fig4}(a) are the DFT-computed dHvA frequencies corresponding to the three Fermi pockets. We find excellent agreement in the angular dependence.
The minimum in $F_{N}$ in the $ac$ sweep occurs when the field is oriented along $c$, \textit{i.e.} along the length of the needle pocket (Fig. \ref{Fig4}(c)), and the maximum appears when the field is aligned along $a$, perpendicular to the needle pocket. $F_{N}$ is independent of angle in the $ab$ sweep because of the rotational symmetry of the needle pocket.
$F_{T1}$ and $F_{T2}$, the frequencies arising from the tube pocket, behave similarly, diverging as the field is rotated from $c$ to $a$. Because the tube is continuous in the $c$ direction and has infinite extreme cross section perpendicular to the $ab$ plane (Fig. \ref{Fig4}(c)), it is expected that the frequency corresponding to the tube pocket can not be seen when the field rotates within the $ab$ plane. This is indeed what we have observed in the right panel of Fig. \ref{Fig4}(a).
The multiplicity and lower symmetry of the diamond pocket create the more complex $F_D$ profiles in both $ac$ and $ab$ sweeps.

Despite excellent agreement in angular dependence, a scaling factor of 0.7 is needed to maximize the quantitative similarity between the experimental and computed values of the extreme cross sections. Small quantitative deviations can occur due to the approximate nature of DFT, particularly when computing flat bands close to the Fermi level \cite{skeaf}.
Furthermore, in a previous study of \ce{NbAs2}, a similar deviation was observed and attributed to As vacancies or imperfections that may lead to subtle relative shifts in bands \cite{NbAs2}.
Crystalline defects, despite being low in \ce{SrAg4Sb2}\cite{stoyko}, could contribute to the difference here since even small shifts can lead to sizable changes in small pockets. With these factors in mind, we conclude that the essential features of the Fermi surface are correctly described by our DFT calculations.

Based on our QO data and the shape of Fermi surface determined from DFT, we can estimate the hole carrier densities associated with the tube pocket and the needle pocket. For the tube pocket, according to the Onsager relation and assuming circular cross sections, the Fermi wave vector $k_F = \sqrt{2eF/\hbar}$ are $k_{T1} =0.11(2)~\si{\AA^{-1}}$ and $k_{T2}=0.16(1) ~\si{\AA^{-1}}$ with the Fermi velocities $v_F = \hbar k_F/m^*$ valued at $v_{{T1}}=4.7(9) \times 10^5~\si{m/s}$ and $v_{{T2}}=8.0(9) \times 10^5~\si{m/s}$, where values $F_{T1} = 420(8)$ T, $F_{T2} = 790(7)$ T, $m^*_{{T1}}=0.27(1)m_e$ and $m^*_{{T2}}=0.23(2)m_e$ are used. Approximating the tube as a cylinder with a radius $k_T=(k_{T1}+k_{T2})/2$ and a height $2\pi/c$, the carrier density $ 2\pi k_T^2 (2\pi/c)/(8\pi^3)$ is estimated to be $p_{{T}}=1.3(3) \times 10^{26}$ m$^{-3}$. We can calculate the carrier density associated with the needle pocket by approximating it as a prolate spheroid. The Fermi wave vectors are $k_{N1} = 0.04(1)~\si{\AA}$ along the minor axis and $k_{N2} = 0.10(2) ~\si{\AA}$ along the major axis with the Fermi velocity of $v_{N1} = 5(1) \times 10^5~\si{m /s}$ and $v_{N2} = 3.3(2) \times 10^5~\si{m/s}$, where $F_{N2} = 340(9)$ T, $F_{N1} = 60(3)$ T, $m^*_{{N2}}=0.35(2)m_e$, and $m^*_{{N1}}=0.097(1)m_e$ are used. The carrier density $2(4 \pi k_{N1}^2 k_{N2}/3)/(8\pi^3)$ is thus estimated as $p_N =0.06(2) \times 10^{26} $ m$^{-1}$ and the total hole carrier density obtained from QO is $p_{\rm{QO}} =1.36 \times 10^{26} $ m$^{-3}$, this rough estimation agrees with $p_{\rm{Hall}}= 3.53 \times 10^{26} $ m$^{-3}$ reasonably well.


\begin{table}
\setlength{\tabcolsep}{10pt}
\renewcommand{\arraystretch}{1.5}
\caption{Two possible sets of topological invariants mapped from the symmetry indicator $\mathbb{Z}_{2,2,2,4}$ = \{1, 1, 1, 2\} for \ce{SrAg4Sb2} \cite{tqc, catalog, catalog-2}.  $m^{\bar{2}10}_{(2)}$ denotes the mirror plane ($\bar{2}10$) in the conventional cell. $g^{\bar{2}10}_{\frac{1}{6}\frac{1}{3}\frac{1}{3}}$ denotes the ($\bar{2}10$) glide plane with glide plane vector ($\frac{1}{6}\frac{1}{3}\frac{1}{3}$). $2^{100}$ is the two-fold rotational axis along the [100] direction. $i$ is the inversion center. $2^{100}_1$ is the screw axis. For a detailed guide of this Table, please refer to Ref. \cite{TCIinvariants}.}
\label{invariants}
\centering
\begin{tabular}{cccccc}
\hline\hline
Weak & $m^{\bar{2}10}_{(2)}$ & $g^{\bar{2}10}_{\frac{1}{6}\frac{1}{3}\frac{1}{3}}$ & $2^{100}$ & $i$ & $2^{100}_1$              \\
\hline
111 & 0 & 1 & 1 & 1 & 1 \\
111 & 2 & 0 & 0 & 1 & 0 \\
\hline\hline
\end{tabular}
\end{table}

\section{Discussion}
The excellent agreement between experiments and DFT calculations suggests the latter correctly describe the band structure of \ce{SrAg4Sb2}. Now let us take a closer look at the electronic band structure of \ce{SrAg4Sb2}, as shown in Fig. \ref{Fig4}(b) and supplemental Fig. S1 (a). Most of the bands near $E_{F}$ are dominated by Sb orbitals with contributions from the Ag orbitals, which is consistent with Sb being an anion with filled $s$ and $p$ orbitals. Above $E_{F}$, the contributions of Sb are highly $k$-dependent. Meanwhile, Sr, a closed-shell cation with little covalency, contributes very little to the bands close to $E_{F}$. Figure \ref{Fig4}(d) shows the computed Fermi surface of \ce{SrAg4Sb2} with projections onto Sb and Ag$^{\rm{trig}}$ orbitals. Sb 5$p_{x,y}$ and Ag$^{\rm{trig}}$ 4$d_{xy,x^2-y^2}$ orbitals dominate the body of the needle pocket (near T point), while the tip of the needle pockets (near $\Gamma$ point) is mainly composed of Sb 5$s$ and Ag$^{\rm{trig}}$ 5$s$ orbitals. The needle pocket thus arises predominantly due to orbital interactions in the $xy$ plane, although near the tips the pocket has more $s$ character. The tube pocket is predominantly composed of Sb 5$p_{x,y}$ and Ag$^{\rm{trig}}$ 4$d_{xy,x^2-y^2}$ orbitals, like the body of the needle pocket. The diamond pocket, on the other hand, has stronger \ce{Ag^{tet}} character. The quasi-cylindrical shapes of the tube pocket and the needle pocket are consistent with the fact that they arise primarily from in-plane interactions in the layered structure of \ce{SrAg4Sb2}, while the diamond pockets likely arise from more isotropic interactions. The compositions also give us a chemical intuition for the charge transfer processes at work in \ce{SrAg4Sb2} - the needle and tube hole pockets are primarily composed of Sb, a formal 3- anion, and the diamond electron pockets are primarily composed of \ce{Ag^{tet}}, a formal 1+ cation, so charge transfer of electrons from Sb to \ce{Ag^{tet}} occurs. 

The most important feature in the band structure is an avoided crossing between the two bands marked by a dashed oval slightly above $E_F$ along the $\Gamma$-T line (Fig. \ref{Fig4}(b)), which lies in the mirror plane.
It has been previously shown to be part of a band inversion, a significant feature for topological materials \cite{morgan_SrAg4Sb2_bonding}.
To characterize the symmetry properties of the band inversion we have used the irvsp code \cite{irvsp}. $\Gamma$ and T both have $D_{3d}$ point symmetry, the highest possible point symmetry in the $R\bar{3}m$ space group.
Since SOC is required to form the band inversion we must describe the band symmetries with double groups, rather than the usual point groups.
At $\Gamma$ the needle band transforms as $E_{\frac{1}{2}g}$ and the tube band transforms as $E_{\frac{1}{2}u}$, while at T (in the inverted state) they are the other way around. 
Points between $\Gamma$ and T have $C_{3v}$ point symmetry, and both bands transform as $E_{\frac{1}{2}}$.
Since the bands have the same symmetry at the intermediate points, they mix and avoid each other, rather than crossing, which creates the band inversion. The band inversion leads to the $k$-dependent changes in orbital composition shown in Fig. \ref{Fig4}(d).

To check where the band inversion occurs, we projected the band structures onto the atomic orbitals.
The band inversion is shown in detail in the projected band structures along the $\Gamma$-T-$\Gamma$ path in Fig. \ref{BandPMOs}, where the projections are onto Ag 5$s$ and Ag$^{\rm{trig}}$ 4$d_{xy,x^2-y^2}$. The left column (fig. \ref{BandPMOs} (a) and (c)) show the bands when they are spin orbit decoupled (SOD) and the right column (fig.\ref{BandPMOs} (b) and (d)) illustrates the changes to the bands when SOC is taken in to account. 
In the SOD regime, the lower band, which is below $E_F$ at $\Gamma$, forms the needle pocket, and the upper band forms the tube and diamond pocket. When SOC is activated, the upper band splits, resulting in a decrease and increase in energy of the tube and diamond bands, respectively. This leads to a distinct band inversion centered at T, formed between the tube and needle bands.

Materials in the space group $R\bar{3}m$ are characterized by four symmetry-based indicators, $\mathbb{Z}_{2,2,2,4}$ = \{1, 1, 1, 2\} \cite{TCIinvariants, tqc}. Odd values associated with $\mathbb{Z}_{4}$ suggests a strong TI, however, for the case of \ce{SrAg4Sb2}, $\mathbb{Z}_{4}$ = 2, thus excluding the possibility of being a strong TI. To further comprehend this, we repeated the symmetry analysis for the SOD electronic structure and found that the symmetry relationships between the bands transform as different irreps at $\Gamma$ and T but the same at all intermediate points. SOC therefore creates the band inversion by influencing the band energies but not their symmetries. The symmetry consequences of SOC are a necessary ingredient for a time-reversal-symmetry protected topological insulator but not for a TCI \cite{TIs colloquium}, so this analysis further supports the claim that \ce{SrAg4Sb2} is a TCI with SOC.

By mapping the symmetry-based indicator to topological invariants, two possible sets of topological invariants exist, as shown in table \ref{invariants}. The first set states that the mirror Chern number $C_m=1$, while the hourglass invariant $\delta_h=1$, the rotation invariant $\delta_r=1$, the inversion invariant $\delta_i=1$ and the screw invariant $\delta_s=1$. This set of invariants indicate 2D Dirac surface states are protected by two-fold rotational symmetry and exist on the ($\bar{2}10$), (110), and (1$\bar{2}0$) planes; 2D Dirac surface states that are protected by glide symmetry occur on the $ab$ plane. The second set states that the mirror Chern number $C_m=2$ and the inversion invariant $\delta_i=1$. In this case, 2D Dirac surface states only exist on the $ab$ plane and are protected by mirror symmetry. From the symmetry-based indicator, it is unclear which set \ce{SrAg4Sb2} takes unless $C_m$ is calculated. For example, in the proposed TCI candidate Bi, the symmetry-based indicator suggests it either has a non-zero mirror Chern number $C_m(1\bar{1}0)=2$ or a non-zero rotation invariant $\delta_r(1\bar{1}0)=1$. And further calculations suggest the latter is the case in the real material \cite{Bi}.

Last but not least, although \ce{SrAg4Sb2} is topologically categorized as a TCI, it is an electronic conductor, having a bulk Fermi surface. 
This aspect of \ce{SrAg4Sb2} is discussed in previous work and coined the term ``semimetallic topological insulator''\cite{morgan_SrAg4Sb2_bonding}.
This property is not unique or new; the same issue was noted for the Bi-Sb alloy system in a landmark paper on $\mathbb{Z}_2$ time-reversal-symmetry protected topological insulators \cite{TIs inversion sym} and has also been discussed for \ce{PbTaSe2} which is a superconducting $\mathbb{Z}_2$ topological insulator, featuring a continuous gap with multiple band crossings rather than a global gap at the Fermi level \cite{PbTaSe2}.

\section{Conclusion}
In summary, we have grown high quality single crystals of the TCI candidate \ce{SrAg4Sb2} and investigated its magnetotransport properties and 3D fermiology. From our magnetotransport measurements, we find a moderately large MR of 700$\%$. From the two-band model fitting we show holes and electrons to posses similar concentrations at low temperatures, suggesting that this is a compensated semimetal. Quantum oscillations are observed in both Hall and magnetic torque measurements. The temperature dependence of the dHvA oscillations reveals small effective masses associated with the Fermi pockets. The angle dependence of the dHvA data shows great agreement with the DFT calculations. Through the analysis of dHvA oscillations and comparison with first-principles calculations, we have demonstrated \ce{SrAg4Sb2} to have a band inversion centered around the T point and contain one needle hole pocket centered at the T point, one tube hole pocket centered at the $\Gamma$ point and a diamond pocket at the F point. Two sets of topological invariants are possible for \ce{SrAg4Sb2}, suggesting that it is a TCI candidate with the 2D Dirac surface states either on the $ab$ planes or on both $ab$ planes and mirror planes, which are all as-grown single crystal surfaces. Further angle-resolved photoemission spectroscopy measurements and theoretical works are urged to confirm its topological properties.

\section*{Acknowledgments}
 Work at UCLA was supported by the U.S. Department of Energy (DOE), Office of Science, Office of Basic Energy Sciences under Award Number DE-SC0021117 to N.N., and the Brown Science Foundation award 1168 to A.N.A.

\medskip

\bibliographystyle{apsrev4-1}
\bibliography{SbMBT}

\end{document}